\documentclass[a4paper,11pt]{article}

\usepackage{contribution}
\usepackage{multirow}

\newcommand{\weblink}[2][]{%
    \ifthenelse{\equal{#1}{}}%
    {\textnormal{\url{#2}}}%
    {\textnormal{\href{#2}{#1}}}%
}

\def\beq{\begin{equation}}
\def\eeq#1{\label{#1}\end{equation}}
\def\eeqn{\end{equation}}

\def\beqa{\begin{eqnarray}}
\def\eeqa#1{\label{#1}\end{eqnarray}}
\def\eeqan{\end{eqnarray}}

\let\bar=\overbar

\def\Dslash{\not{\hbox{\kern-4pt $D$}}}
\def\dslash{\not{\hbox{\kern-2pt $\del$}}}

\def\msb{{\bar{\ssstyle M \kern -1pt S}}}

\newcommand{\contribution}[7][]{%
  \clearpage
  \thispagestyle{plain}
  \ifthenelse{\equal{#1}{}}
  {\hypersetup{pdftitle={#2}}}
  {\hypersetup{pdftitle={#1}}}
  \hypersetup{pdfauthor={{#3} {#4}}}
  {\centering\normalfont\LARGE\bfseries\sffamily #2 \par\nobreak}
  \lhead{}
  \chead{%
    \textit{\footnotesize XIV International Conference on Hadron Spectroscopy
      (\weblink[\textit{hadron2011}]{http://www.hadron2011.de}), 13-17 June 2011, Munich, Germany}%
  }
  \rhead{}
  \bigskip
  \begin{center}
    {#3} {#4}\ifthenelse{\equal{#6}{}}{}{\footnote{\weblink[#6]{mailto:#6}}}
    \ifthenelse{\equal{#7}{}}{}{#7} \\
    \textit{#5}
  \end{center}
  \bigskip
}

\renewcommand{\abstract}[1]{%
  \begin{center}
    \begin{minipage}{0.85\textwidth}
      \begin{footnotesize}
        #1
      \end{footnotesize}
    \end{minipage}
  \end{center}
  \bigskip
}

\begin{document}

%
%
%
%
%
{  

\makeatletter
\@ifundefined{c@affiliation}%
{\newcounter{affiliation}}{}%
\makeatother
\newcommand{\affiliation}[2][]{\setcounter{affiliation}{#2}%
  \ensuremath{{^{\alph{affiliation}}}\text{#1}}}
%


\contribution[]
{The nature of the lightest scalar meson, its $N_c$ behaviour and semi-local duality}
{J.R. }{Pelaez}  
{\affiliation[\footnotesize Dept. de F\'isica Te\'orica II,
Facultad de Ciencias F\'isicas, Universidad
Complutense de Madrid, E-28040, Madrid, Spain]{1} \\
 \affiliation[\footnotesize Theory Center, Thomas Jefferson National Accelerator Facility, 
12000 Jefferson Av., Newport News, VA 23606, U.S.A]{2} \\
 \affiliation[\footnotesize Physics Division, Argonne National Laboratory, Argonne, Illinois 60439,
    U.S.A.]{3}}
{}
{\!\!$^,\affiliation{1}$, M.R. Pennington {\affiliation{2}}, J. Ruiz de Elvira\affiliation{1} and D.J. Wilson\affiliation{3}}

%

\abstract{%
 
One-loop unitarized Chiral Perturbation Theory (UChPT) calculations, suggest a different $N_c$ behaviour for
the $\sigma$ or $f_0$(600) and $\rho$(770) mesons: while the $\rho$ meson becomes
narrower with $N_c$, as expected for a $\bar{q}q$ meson, the $f_0$(600)
contribution to the total cross section below 1 GeV becomes less and less
important.  
Here we review our recent work \cite{RuizdeElvira:2010cs}
where we have shown, by means of finite energy sum rules,
that a different $N_c$ behaviour for these resonances may lead to a conflict 
with semi-local duality for large $N_c$, since local duality requires a cancellation between the $f_0$(600)
and $\rho$(770) amplitudes.
However, UChPT calculations also suggest a subdominant
$\bar{q}q$ component for the $f_0$(600) with a mass above 1 GeV and this can restore semi-local duality, as we show. 

}
%

\section{Introduction}
QCD perturbative calculations are not applicable to the 
longstanding \cite{Jaffe:1976ig} controversy on the non-$\bar q{q}$
nature of light scalar mesons.
However, the QCD 1/$N_c$ expansion \cite{'tHooft:1973jz} 
allows for a clear identification of a $\bar{q}q$ resonance,
since it becomes a bound state, whose width behaves O(1/$N_c$), 
and its mass as O(1). In addition, 
in this low energy region one can use Chiral
Perturbation Theory (ChPT) \cite{Weinberg:1978kz,ChPT} to obtain a model independent description of the dynamics of pions, kaons and etas, which are the pseudo-Goldstone
Bosons of the QCD spontaneous breaking of Chiral Symmetry.
Lately, by combining the $1/N_c$ expansion of ChPT with dispersion theory, it has been possible to
study the nature of light resonances thus generated in meson-meson scattering 
\cite{Pelaez:2003dy,Pelaez:2006nj}.

Let us recall that the ChPT Lagrangian 
is built as a low energy expansion
respecting all QCD symmetries, and using only pseudo-Goldstone boson fields. The small masses of the three lightest quarks
can be treated perturbatively and thus ChPT becomes a series in momenta and meson masses, generically
$O(p^2/\Lambda^2)$. Apart from these masses and $f_\pi$ ---the pion decay constant, which sets the scale 
$\Lambda \equiv 4\pi f_\pi$ ---there are no free parameters at leading order.
The chiral expansion is renormalized order by order by absorbing loop divergences
in the coefficients of higher order counterterms, called low energy constants (LECs), whose values 
depend on the underlying QCD dynamics, and have to be determined from experiment. 
However, the leading $1/N_c$ dependence of the LECs is known and model 
independent \cite{ChPT}, thus
allowing us to study the $N_c$ dependence of low energy hadronic observables.

\subsection{Unitarization and dispersion theory}
Unitarity implies that, for physical values of $s$, partial
waves $t^{IJ}$ of definite isospin $I$ and angular momenta $J$ for \textit{elastic} meson-meson
scattering satisfy:
\begin{equation}
  {\rm Im}\;t^{IJ}=\sigma
  |t^{IJ}|^2\;\;\;\Rightarrow\;\;\;{\rm Im}\frac{1}{t^{IJ}}=-\sigma
\label{unit}
\end{equation}
where $\sigma = 2p/\sqrt{s}$, and $p$ is the CM momentum. Note that
unitarity implies that $|t^{IJ} | \leq 1/\sigma$, and
 typically elastic resonances are characterized
by the saturation of this bound.
However, ChPT partial waves are a low energy expansion
$t \simeq t_2+t_4+t_6+\cdots$, (we will drop
the $IJ$ indices for simplicity.) where $t_{2k} \equiv O(p/(4\pi f_\pi))^{2k}$, 
and thus they cannot satisfy
unitarity exactly, but just perturbatively, i.e: $\mathrm{Im}\,t_2 = 0,\;\;\; \mathrm{Im}\,t_4 = \sigma t_2^2,\;\;\;\; etc\, ...$

The elastic Inverse Amplitude Method (IAM) \cite{Truong:1988zp,Dobado:1992ha} uses ChPT to evaluate the subtraction 
constants and the left cut of a dispersion relation for the inverse
of the partial wave. The elastic right cut is calculated exactly with Eq.\eqref{unit}---thus ensuring elastic unitarity. Note that the IAM is 
derived only from elastic 
unitarity, analyticity in the form of a dispersion
relation, and ChPT, which is only used at low energies. Remarkably, the IAM can be rewritten as a simple
algebraic formula in terms of ChPT amplitudes, but it 
satisfies exact elastic unitarity, at low energies recovers the chiral expansion up to the initially given order,
 and reproduces meson-meson
scattering data up to energies $\sim$ 1 GeV. This is done with values
of the LECs which are fairly compatible with the values obtained within
standard ChPT. Since it is derived from a dispersion relation, it can be 
analytically continued into the second Riemann sheet where, within 
the SU(2) ChPT formalism that we use here, we find the
poles associated to the $\rho(770)$ and $f_0(600)$ resonances.
Hence, we can study,
without any \textit{a priori} assumption, the nature of the $\rho(770)$ and $f_0(600)$ from
first principles and QCD.

\subsection{The 1/$N_c$ expansion}
 For our purposes \cite{RuizdeElvira:2010cs}, the relevant
observation is that the leading 1/$N_c$ behaviour of the ChPT LECs is
known. Thus, in order to obtain
the leading $N_c$
behaviour of the resonances generated with the IAM, we just have to rescale
$f_\pi\rightarrow f_\pi \sqrt{N_c/3}$, the one-loop LECs as
$l^r_{i}\rightarrow l_i^r N_c/3$ and 
the two loop ones as $r_{i}\rightarrow r_i (N_c/3)^2$.

This procedure \cite{Pelaez:2003dy} was first applied to the one-loop SU(3) ChPT amplitudes, 
and the result was that the light vector resonances, 
as for example the $\rho$(770), followed the
expected behaviour of $\bar{q}q$  states remarkably well, as we show in the left panel of Fig.\ref{fig:poles}.
In contrast, as seen on the right panel of  Fig.\ref{fig:poles} the
$f_0$(600) behavior is at odds with that of $\bar{q}q$ states.
It follows that the dominant component of the $f_0$(600) ( and the other
members of the lightest scalar nonet) does not have a $\bar{q}q$ nature.

   \begin{figure}[t]
     \centering
     \includegraphics[width=0.47\textwidth]{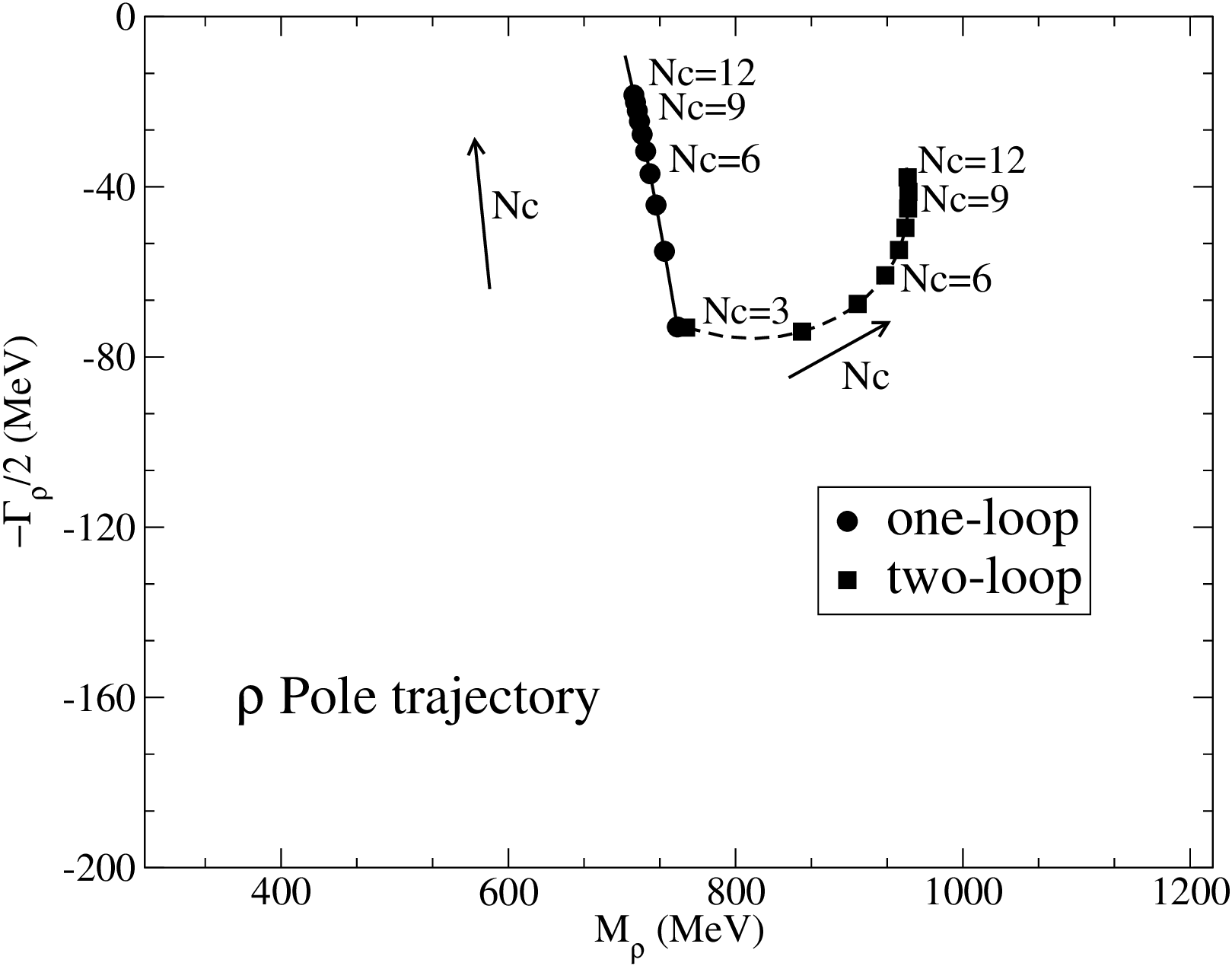}
     \includegraphics[width=0.47\textwidth]{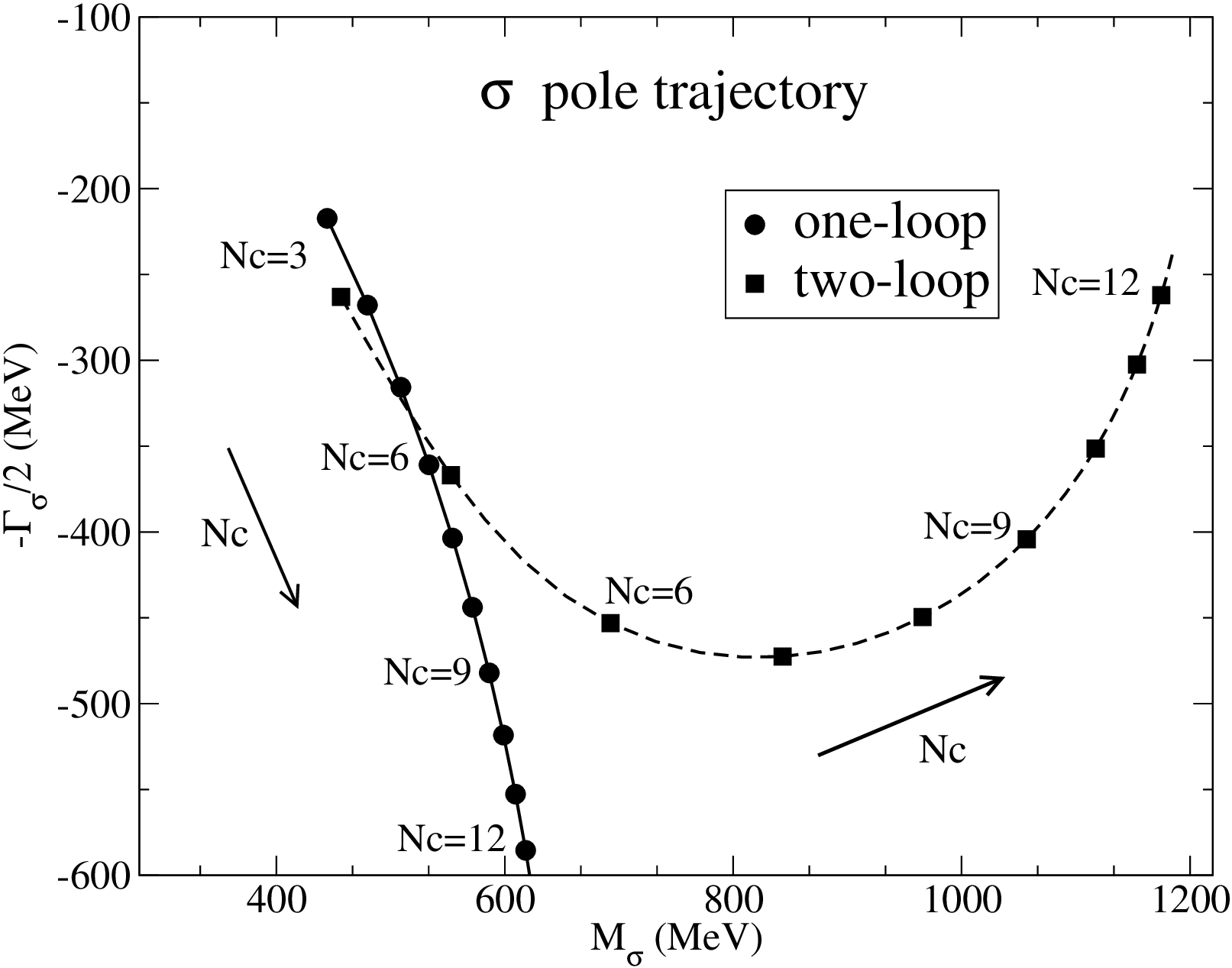}
     \caption{Left, the $\rho$ meson behaves as a $\bar{q}q$ state. Center: At O($p^4$)   the  $\sigma$ does not. Right: At O(p$^6$) emerges a subleading $\bar{q}q$
   component at 1.2 GeV}
 \label{fig:poles}
   \end{figure}

Of course, these results have some uncertainty, particularly on the renormalization scale
where the $N_c$ scale is to be applied, which was studied also in \cite{Pelaez:2003dy}. However,
the general conclusions are rather robust: whereas vector mesons behave predominantly as a $\bar{q}q$,
the scalar ones do not, at least for $N_c$ not far from 3, which is where the IAM can be applied
(not at $N_c\rightarrow\infty$, see \cite{Pelaez:2010er})

The two-loop SU(2) ChPT amplitude analysis  \cite{Pelaez:2006nj} 
showed that when the $\rho$(770) is to follow its expected $\bar{q}q$
behaviour, the $f_0$(600) still did not follow a $\bar{q}q$ behavior at first,
and its pole moved away from the 400-600 MeV region of the real axis.
However, for $N_c\sim$ 8, the $f_0$(600) started behaving
as a $\bar{q}q$, see Fig. 1 (Right). 
In conclusion, the two-loop IAM
 confirms once again that the $f_0$(600) does not
behave predominantly as a $\bar{q}q$ state, but suggests the existence of a
subdominant $\bar{q}q$ component
that originates at a mass of $\simeq$ 1.2 GeV, which is approximately
twice that of the physical  $\sigma$ at $N_c=3$. This seems to support
models like \cite{vanBeveren:1986ea} that have indeed suggested 
a non-$\bar qq$ nonet below 1 GeV and an additional $\bar qq$ one
above.

\section{Semi-local duality}
A well known feature of the real world ($N_c$ = 3) is that of ``local duality''. At low energies the scattering amplitude is well represented by a sum of resonances
(with a background), but as the energy increases the resonances (having more phase space for decay)
become wider and increasingly overlap. This overlap generates a smooth Regge 
behaviour described by a small number of crossed channel Regge exchanges. 
Indeed, detailed studies of meson-baryon
scattering show that the sum of resonance contributions at all energies ``averages'' 
 the higher energy Regge behaviour. Thus, s-channel resonances
are related to Regge exchanges in the t-channel and are ``dual'' to each other: one uses one
or the other. 

Regge exchanges are also built from $\bar{q}q$ and multi-quark
contributions. However, in the isospin 2 
$\pi\pi$ scattering channel there are no $\bar{q}q$  resonances, 
and so the Regge exchanges with these quantum number must involve multi-quark components. 
Data teach us that even
at $N_c$ = 3 these components are suppressed 
compared to the dominant $\bar{q}q$ exchanges. hence, semi-local
duality means that in $\pi^{+}\pi^{-} \rightarrow \pi^{-}\pi^{+}$, which is an isospin 2 process, 
the low energy resonances must have contributions to
the cross-section that ``on the average'' cancel. In particular, 
using the crossing relations the I=2 t-channel amplitude can be recast as
a function of s-channel amplitudes: 
\begin{equation}
\mathrm{Im}\,A^{t2}(s,t)=\frac{1}{3}\mathrm{Im}\,A^{s0}(s,t)-\frac{1}{2}\mathrm{Im}\,A^{s1}(s,t)+\frac{1}{6}\mathrm{Im}\,A^{s2}(s,t),
\end{equation}
but since $A^{s2}$ is repulsive and small, the strong cancellation occurs between 
$A^{s0}$ and $A^{s1}$. However, these channels are saturated at low energies by 
the  $f_0$(600) and $\rho$(770) resonances, respectively.
Hence, semi-local duality requires 
the contribution of these two resonances to cancel
``on average'' in keeping with I = 2 exchange in the
t-channel. 
This ``on the average cancellation'' is properly defined via Finite Energy Sum
Rules:   
\begin{equation}\label{FESR}
F(t)^{21}_n= \frac{\int_{\nu_{th}}^{\nu_{\mathrm{max}}}{d\nu\;\mathrm{Im}\, A^{t2}(s,t)/\nu^n}}{\int_{\nu_{th}}^{\nu_{\mathrm{max}}}{d\nu\;\mathrm{Im}\,A^{t1}(s,t)/\nu^n}},\;\;\;\nu=(s-u)/2.
\end{equation}

Semi-local duality between Regge and resonance contributions teaches us that
on the ``average'' and at least over one resonance tower, we have:
\begin{equation}\label{ReggeLocal}
  \int_{\nu_{\mathrm{th}}}^{\nu_{\mathrm{max}}}{d\nu\;\nu^{-n}\mathrm{Im}\,A^{t2}(s,t)_{\mathrm{Data}}}\sim \int_{\nu_{\mathrm{th}}}^{\nu_{\mathrm{max}}}{d\nu\;\nu^{-n}\mathrm{Im}\,A^{t2}(s,t)_{\mathrm{Regge}}}
\end{equation}

where Regge amplitudes are given as usual for $\alpha' \nu \gg 1$ by
  \begin{equation}\label{Regge}
{\rm Im} A^{tI}(\nu,t) \simeq \sum_R \,
\beta_R(t)\,[\alpha'\,\nu]^{\alpha_R(t)}
  \end{equation}
(see [1] for its low energy extrapolation), and  where
$\alpha_R(t)$ denote the Regge trajectories with the
 appropriate $t$-channel quantum numbers, 
 $\beta_R(t)$ their Regge couplings and $\alpha'$ 
 is the universal slope of the $q \bar q$ meson
 trajectories ($\sim 0.9$ GeV$^{-2}$). 
 For the $I=0$ exchange the dominant trajectories are the Pomeron and the $f_2$
 with intercepts close to 1 and 0.5 respectively, while the $I=1$ $\rho$ exchange is degenerate with the $f_2$.
 For the exotic $I=2$ channel with its leading Regge exchange being a $\rho-\rho$ cut, we expect its intercept to be much smaller than that of the $\rho$, 
 and its couplings to be correspondingly smaller.
  Therefore using Eqs.\eqref{ReggeLocal} and \eqref{Regge}, local duality implies
 that $|F(t)^{21}_n| \ll 1$.

We can now use the IAM to check the  $N_c$ dependence of $\pi\pi$
scattering amplitudes. In particular the I=2 s-channel amplitude remains
repulsive with $N_c$, and still there is no resonance exchange. Therefore semi-local
duality implies that the $t$-channel $I=2$ Regge exchange should continue to
be suppressed as $N_c$ increases. Since the Regge trajectories do not depend on $N_c$, still
$|F(t)^{21}_n| \ll 1$ when increasing $N_c$ due to a strong cancellation between the
$\rho$(770) and the $f_0$(600) which would not occur if the $f_0$(600) disappeared completely from the spectrum.

\section{Results}

\begin{table}
\centering
    \begin{tabular}{c|c||c|c|c}
      $\nu_{max}$ & 400 GeV$^2$ &2.5 GeV$^2$&2 GeV$^2$&1 GeV$^2$ \\\hline 
      $F^{21}_1$ &0.021 $\pm$ 0.016 & 0.180 $\pm$ 0.066 &0.199 $\pm$ 0.089 & -0.320 $\pm$ 0.007\\\hline
      $F^{21}_2$ &0.057 $\pm$ 0.024 & 0.068 $\pm$ 0.024 &0.063 $\pm$ 0.025 & -0.115 $\pm$ 0.013\\\hline
      $F^{21}_3$ &0.249 $\pm$ 0.021 & 0.257 $\pm$ 0.022 &0.259 $\pm$ 0.022 & 0.221  $\pm$ 0.021\\\hline
    \end{tabular}
    \caption{Values of
        the ratio F$^{21}_n$  using the KPY parametrization and different cutoffs.
      All $F^{21}_n$ ratios for a 20 GeV cutoff turn out very small, of the order 1:50 or 1:15 . 
      However, we see that they are only 1:4 or 1:5 when $s_{max}$ is still
      $\sim$2 GeV$^2$}\label{table 1}
\end{table}

Using the IAM we can study the behaviour of Eq. \eqref{FESR} with $N_c$ \cite{RuizdeElvira:2010cs}, and
then check if $|F(t)^{21}_n| \ll 1$ when increasing the number of colors. 
However, the IAM is only valid in the low energy region, and we have to
check the influence of the high energy part on this cancellation.
For this reason, in Table \ref{table 1} we 
first calculate the value of the FESR for
different \textit{cutoffs} using the 
$\pi\pi$ KPY \cite{Kaminski:2006qe} data parametrization  
 as input. We thus check
that local duality is satisfied for $N_c$=3 since $|F(t)^{21}_n|
\ll 1$, and at least for n=2,3, the main contribution to the FESR
suppression occurs below 1 GeV, where we can apply
the IAM. Therefore, we can use the IAM to study local duality, and check the FESR
suppression with $N_c$. 
In evaluating the amplitudes $\mathrm{Im}A^{sI}$, we represent these by a sum of s-channel
partial waves, so that:
\begin{equation}
  \mathrm{Im}\,A^{sI}(s,t)=\sum_{J}{(2J+1)\mathrm{Im}\,t^{IJ}(s)P_J(\cos(\theta_s))}.
\end{equation}
However, using the IAM only S0, P and S2 waves can be described. It it is
necessary to check the effect of those waves in Eq. \eqref{FESR}. In Table
\ref{table 2} we show how the influence of higher waves is around
10$\%$, and that the IAM predicts correctly the FESR suppression. 
\begin{table}
\centering
    \begin{tabular}{c|c|c|c}\hline
      \multicolumn{4}{c}{$\nu_{max}$=1 GeV$^2$} \\ \hline
      & KPY08& 1 loop UChPT  & 2 loop UChPT \\\hline\hline 
      $F^{21}_1$ & -0.350 $\pm$ 0.083&-0.355 $\pm$ 0.061& -0.351\\\hline
      $F^{21}_2$ & -0.131 $\pm$ 0.042&-0.157 $\pm$ 0.097& -0.172\\\hline
      $F^{21}_3$ & 0.215 $\pm$ 0.027 &0.175 $\pm$ 0.138 &0.145\\\hline
    \end{tabular}
    \caption{ Comparison between the $F^{21}_n$ at $t=4M_\pi^2$, using only S and P waves with a \textit{cutoff} of 1 GeV$^2$, calculated with data
      parametrizations or our IAM amplitudes.}\label{table 2}
\end{table}

Let us now increase $N_c$:  if 
the $\rho$(770) mass remains constant
and its width becomes narrower, 
but the $f_0$(600) 
contribution to the total cross section below 1 GeV
becomes less and
less important, then the ratios $|F(t)^{21}_n|$ grow and there is a 
conflict with semi-local duality. This is shown in the thin lines
of Figure 2 (Note the gray area above $N_c=30$, where we consider the IAM
merely qualitative). Note however, that this is a generic problem for any model where the $f_0$(600) contribution vanishes, not just the IAM.
However if there is a subdominant $\bar{q}q$
component for the $f_0$(600) with a mass somewhat above 1 GeV, 
as occurs naturally within the two-loop IAM---but also
in a part of the one-loop parameter space---this is enough
to ensure the cancellation with the $\rho$(770) contribution.
The effect is shown by the thick lines of  Figure 2.

\begin{figure}[htb]
  \begin{center}
  \includegraphics[width=0.49\textwidth]{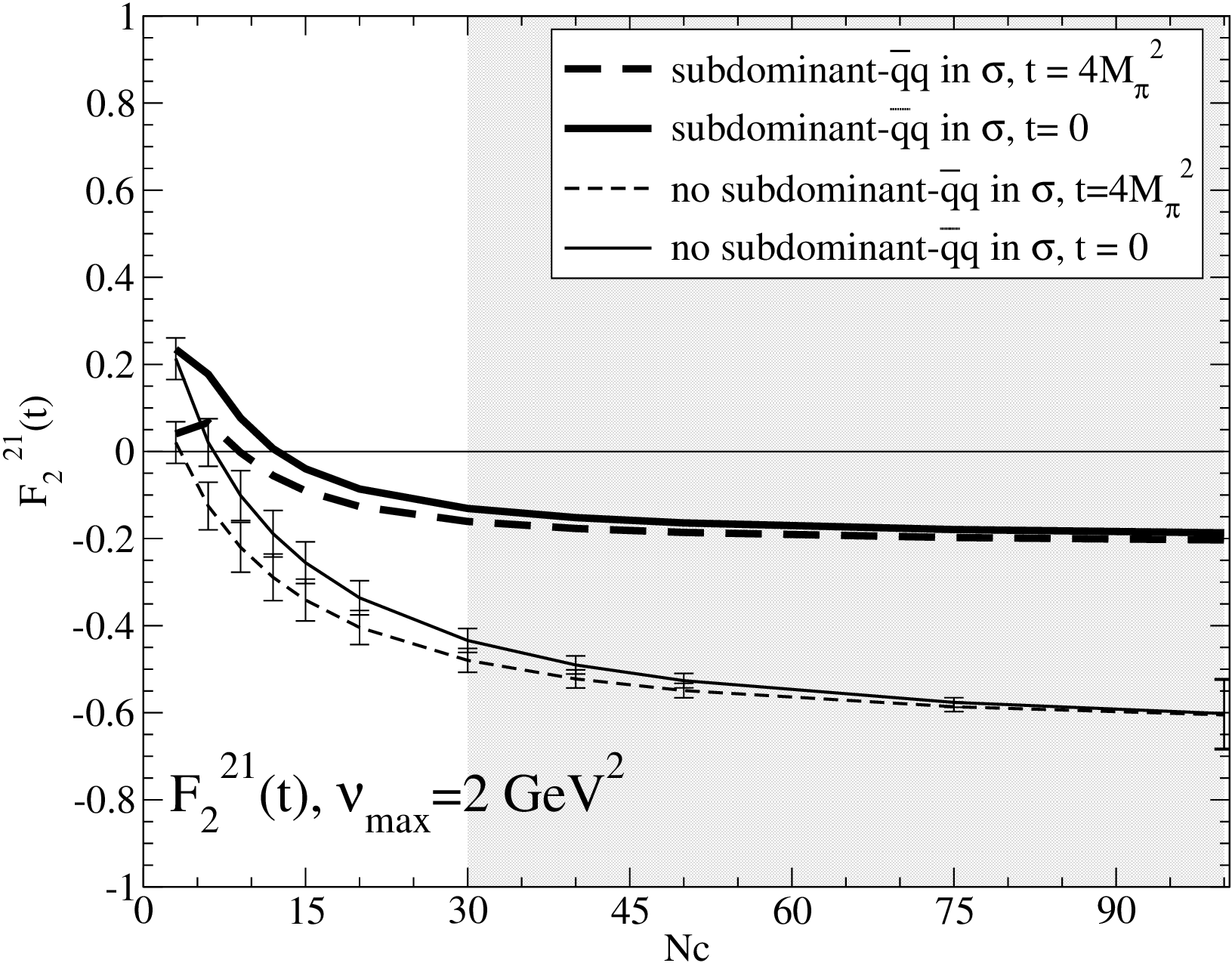}
  \includegraphics[width=0.49\textwidth]{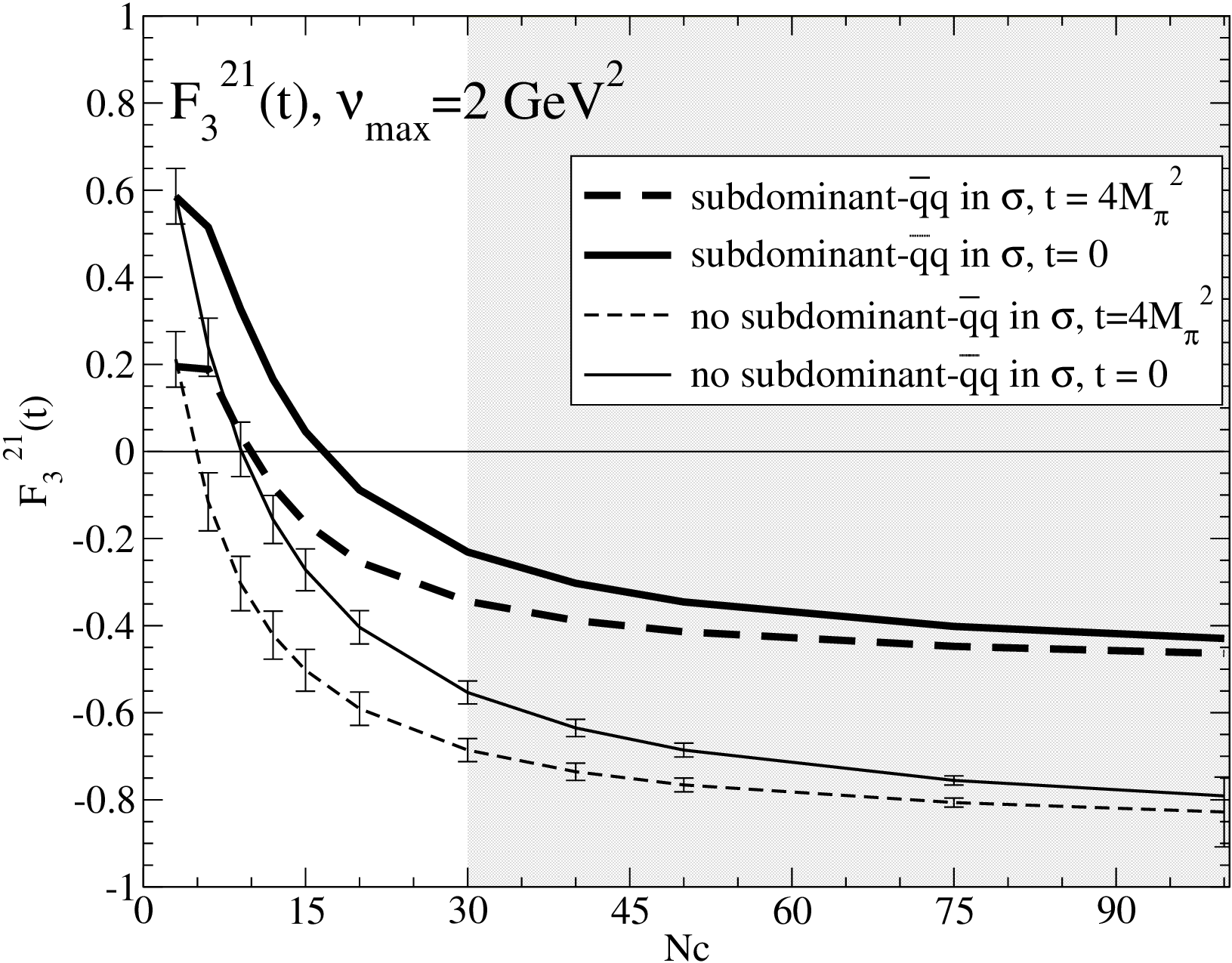}
  \caption{At O($p^4$), solid line, there is no FESR suppression and local duality
    fails as $N_c$ grows. However, at O($p^6$), dashed line, 
the $\sigma$ subleading $\bar{q}q$ component ensures local
    duality even when increasing $N_c$}
\label{Fevolution}
\end{center}
\end{figure}

\section{Conclusions}
The 1/$N_c$ expansion of ChPT unitarized using the IAM shows that the 
$f_0$(600) meson is not predominantly a $\bar{q}q$ state, since the
$\sigma$ amplitude becomes smaller and smaller below 1 GeV as $N_c$ increases.
This different behavior from that of a $q\bar{q}$ state as the $\rho$(770),
leads to a potential conflict with semi-local duality. 
However, unitarized ChPT calculations also suggest a subdominant
$\bar{q}q$ state that emerges somewhat above 1 GeV. 
This subdominant component ensures that semi-local
duality is still satisfied as $N_c$ increases.

\section*{Acknowledgments}
J.R.P. thanks the organicers of Hadron 2011 for support and for creating such a 
stimulating conference.
Work partially supported by Spanish Ministerio de
Educaci\'on y Ciencia research contracts: FPA2007-29115-E,
FPA2008-00592 and FIS2006-03438,
U. Complutense/ Banco Santander grant PR34/07-15875-BSCH and
UCM-BSCH GR58/08 910309. We acknowledge the support
of the European Community-Research Infrastructure
Integrating Activity
``Study of Strongly Interacting Matter''
(acronym HadronPhysics2, Grant Agreement
n. 227431)
under the Seventh Framework Programme of EU.
This paper has in part been authored by Jefferson Science Associates, LLC
 under U.S. DOE contract No. DE-AC05-06OR23177 and supported by the U.S.
DOE, Office of Nuclear Physics, Contract No. DE-AC02-06CH11357.


%

}  


\end{document}